\documentclass[12pt,preprint]{aastex}
\def\lsim{\lower.5ex\hbox{$\; \buildrel < \over \sim \;$}}
\def\gsim{\lower.5ex\hbox{$\; \buildrel > \over \sim \;$}}

\begin{document}

\title{Possible Photometric Evidence of Ejection of Bullet Like Features in the Relativistic Jet source SS433} 

\author{Sandip K.\ Chakrabarti$^{1,2}$,  S. Pal$^{2}$,  A. Nandi$^1$, B.G. Anandarao$^3$, Soumen Mondal$^3$} 

\affil{$^1$S.N. Bose National Center for Basic Sciences, JD-Block, Salt Lake, Kolkata, 700098\\ India\\
$^2$ Centre for Space Physics, P-61 Southend Gardens, Kolkata, 700084, India\\
$^3$  Physical Research Laboratory, Navarangapura, Ahmedabad, 380009, India\\
e-mail: chakraba@bose.res.in, space\_phys@vsnl.com, anuj@bose.res.in, anand@prl.ernet.in, soumen@prl.ernet.in}

\begin{abstract}
SS433 is well-known for its precessing twin jets having
optical bullets inferred through {\it spectroscopic} observation of $H_\alpha$ 
lines. Recently, Chakrabarti et al. (2002) described processes which may be 
operating in accretion disk of SS433 to produce these bullets. In a recent multi-wavelength campaign,
we find sharp rise in intensity in time-scales of few minutes in X-rays, IR and radio waves
through {\it photometric } studies. We interpret them to be 
possible evidence of ejection of bullet-like features from accretion disks.
\end{abstract}


\noindent ASTROPHYS. J. LETTERS, 595, L45, 2003

\section{Introduction}

SS433 is a well studied bright emission line compact system which is ejecting matter
in symmetrically opposite directions at a speed of $v_{jet} \sim 0.26$c.
It has a mass-losing (${\dot M} \sim 10^{-4} M_\odot/$yr) companion
orbiting in $13.1$d. The jets are precessing in $162.15$d around the symmetry axis. The velocity is
remarkably constant to within a percent or so (Margon, 1984; Gies et al. 2002). As a result of the remarkable
constancy in the precession time scales and the velocity, the instantaneous locations of the red and blue-shifted $H_\alpha$
lines are well predicted by the so-called `kinematic-model' (Abell \& Margon, 1979) and the compilation of
twenty years of timing properties (Eikenberry et al. 2001) suggest that kinematic model can explain the general
variation of the red and blue-shifts very well.

A very exciting observation, made within years of the discovery of SS433, suggests that the
passage of the jets is not continuous, but as if through successive and discrete bullet-like entities,
at least in the optical region (Grandi, 1981; Brown et al. 1991). The $H_\alpha$ lines were seen to brighten up
and fade away without changing their red/blue-shifts, indicating the brightened bullets
are radially ejected and do not have any rotational velocity component. Since the bullets
of energy $\sim 10^{35}$ ergs do not change their speed for a considerable time
($\sim 1-2 $ days), Chakrabarti et al (2002) postulated that they must be ejected from accretion
disk itself. They presented a mechanism to produce quasi-regular bullets.
Using results of numerical simulations involving oscillation of shocks in accretion disks, 
they concluded that in the normal circumstances, $50-1000$s interval is 
expected in between the bullet ejection. These bullets would be ejected from X-ray emitting 
region and propagate through optical, infra-red ($\sim 10^{13-14}$cm) and finally 
to radio emitting region at $\gsim 10^{15}$cm (roughly the distance
covered in a day with $v\sim v_{jet}$) or so. Thus
if the object is in a low or quiescence state, each individual bullet flaring and dying away in
a few minutes time scale, should be observable not only in optical wavelengths
(Grandi, 1981; Margon, 1984; Brown et al. 1991; Gies et al. 2002) but also in all the
wavelengths, including X-ray, IR and radio emitting regions.
So far, no such observations of individual bullets has been reported in the literature
and it was necessary to make a multi-wavelength observation at relatively quieter states.

In this {\it Letter}, we present some results of our multi-wavelength studies.
From the long term analysis of radio flares (Bonsignori-Facondi, Padrielli, Montebugnoli 
and Barbieri, 1986; Vermeulen et al. 1993) it is known that in between 
big flares which occupy $\sim 20\%$ of the time the object may go to 
very quiescence state. So, it is likely that one could `catch' these bullets
in action provided observations are carried with very short time resolution. Our multi-wavelength
observations lasted during 25th of Sept., 2002 to 6th of October, 2002 with 
X-ray, infra-red, optical, radio observations made simultaneously on the 27th of September, 2002.
Here, we report only X-ray, infra-red and radio observations of 27th and 29th of September, 2002.  
Optical studies required longer integration times and these results
along with other days of observations would be reported elsewhere (Chakrabarti et al. 2003).

Our main results indicate that there are considerable variations in the timescale of minutes
in all the wavelengths. These may be called micro-flares. When Fourier transform is made, 
some excess power is observed in $2-8$ minutes time scale (often beyond 3$\sigma$ level). 
The X-ray count rate was found to increase by $15-20$\% within a minute. Since the emitting 
regions of X-ray, IR and radio are not well known with absolute certainly, while duration of the
flares last less than a few minutes, we could not prove beyond doubt that 
there are indeed correlations among the micro-flares observed 
in these wavelengths. However variabilities we find are not flicker type or shot noise type in the sense
that the power density spectrum (PDS) is not of $1/f$ type and the duration is not very short (i.e., $<1$s).
We therefore believe that we may have found evidences of bullet ejection through these observations
in other wavelengths.

\section{Observations and Data Reduction}

Radio observation was carried out with Giant Meter Radio Telescope (GMRT) at 1280MHz (bandwidth $16$ MHz)
which has $30$ antennas each of $45$m in diameter spreaded over $25$km 
region (Swarup et al. 1991) near Pune, India along roughly $Y$ 
shaped array. The data is binned at every $16$ seconds. On 27th and 29th of
September, 2002, no. of antennas working were 28 and 13 respectively.
AIPS package was used to reduce the data. 
Bad data was flagged using tasks UVFLG and TVFLG and the standard deviation
at each time bin using UVPLT package was computed. On 27th, 3C48 and
3C286 were used as the flux calibrator while on the 29th only 3C48 was used. 
Generally, the observation condition was very stable.

Infrared observation was made using Physical Research Laboratory (PRL) 1.2m Mt.
Abu infrared telescope equipped with Near-Infrared Camera and Spectrograph 
(NICMOS) having 256 $\times$ 256 HgCdTe detector array cooled to $77$K. The
filters used were standard J ($\lambda$=1.25 $\mu$m, $\Delta\lambda$= 0.30
$\mu$m), H ($\lambda$=1.65 $\mu$m, $\Delta\lambda$= 0.29 $\mu$m) and K$^\prime$
($\lambda$=2.12 $\mu$m, $\Delta\lambda$= 0.36 $\mu$m) bands. The observational
data on 27th of September, 2002 for J and H bands are binned at every 10
seconds while that of K$^\prime$ band is binned at every 20 seconds. The
data reduction was performed using the IRAF software package. 
All the object frames were de-biased, sky-subtracted and
flat-fielded using normalized dome flats. The sky frame was created by usual
practice of median combining of five position dithered images in which the
source was within the NICMOS field of 2$^\prime$ $\times$ 2$^\prime$. At each
dithered position ten frames were taken with each integration time of $10$
seconds. The nearby infrared bright standard star GL748 (Elias et al. 1982)
was used as the calibrator and it was observed for at least 15 minutes at each
filter band. We measured the stellar magnitudes using the aperture photometry
task (APPHOT) in IRAF. Our derived mean JHK$^\prime$ magnitudes on Sept. 27th
are 9.47$\pm$0.02,  8.48$\pm$0.02 and  8.32$\pm$0.02
respectively and the corresponding mean flux densities are 0.261$\pm$0.002,
0.413$\pm$0.003 and 0.305$\pm$0.003 respectively. The magnitudes are converted
to flux density (Jansky) using the zero-magnitude flux scale of Bessell,
Castelli \& Plez (1998). To estimate reddening, we assumed visual extinction
A$_v$ =8.0 (Gies et al. 2002). Using the relation given in Bessell, Castelli
\& Plez (1998) the JHK$^\prime$ extinctions were found to be A$_J$= 2.32, A$_H$
= 1.84 , A$_K$ = 0.88. The de-reddened flux in JHK$^\prime$ are 2.24 Jy, 2.21
Jy and 1.67 Jy respectively. The differential magnitudes are determined using
two brightest stars (std1: J= 12.1, H=10.6, std2: J=12.5, H= 11.1 mag) in
the same frame of the object. The error in individual flux density measurement is usual propagation
error of the observed photometric magnitude. Photometric errors $\epsilon$
are calculated for individual frame of every star and for the subtracted
differential magnitude the final error was calculated as
$\sqrt{\epsilon_1^2 + \epsilon_2^2}$, where $\epsilon_1$ and $\epsilon_2$ are the
error-bars of the individual stars.

X-ray observation was carried out using Proportional Counter Array (PCA) 
aboard RXTE satellite. The data reduction and analysis was performed 
using software (LHEASOFT) FTOOLS 5.1 and XSPEC 11.1. We extract 
light curves from the XTE/PCA Science Data of GoodXenon mode.  We combine 
the two event analyzers (EAs) of 2s readout time to reduce the Good Xenon data
using the perl script {\bf make\_se}. Once {\bf make\_se} script was run on the
Good\_Xenon\_1 and Good\_Xenon\_2 pairs, the resulting file was reduced
as Event files using {\bf seextrct} script to extract light curves. Good time 
intervals were selected to exclude the occultations by the earth and South 
Atlantic Anomaly (SAA) passage and also to ensure the stable pointing. 
We also extracted energy spectra from PCA {\bf Standard2} data in the energy range
$2.5$ - $20.0$ keV (out of five PCUs only data from 0, 2, 3 PCUs are
added together). For each spectrum, we subtracted the background data that
are generated using PCABACKEST v4.0. PCA detector response matrices are
created using PCARSP v7.10.  

\section{Results on short time-scale variabilities}

The observational result of September 27th, 2002 is shown in Fig. 1
with  UT (Day) along the X-axis.
The upper and middle panels show the radio and IR fluxes (uncorrected for reddening) in Jansky
and the lower panel shows X-ray counts per second in $2-20$keV. 
Typical error-bars of the mean-flux measurements 
(standard deviation in each time bin for radio and IR, and squared-root of counts
per binsize for X-rays) which are included in the Figure are:  in radio 
$\sim 1$mJy, in IR  $0.5$mJy and in X-ray $\sim 3$ counts/s.
These observations correspond to an average flux of $10^{-14}$ergs/cm$^2$/s,
$5\times 10^{-10}$ergs/cm$^2$/s and $ 10^{-10}$ergs/cm$^2$/s respectively. In other words,
assuming isotropic emission, at a distance of $3$kpc for the source, the average radio, IR, and X-ray
luminosities are $1.1\times 10^{30}$ ergs/s, $5.5 \times 10^{34}$ergs/s and $10^{35}$ergs/s
respectively. Observations in radio and IR were carried out during
25-30th  September, 2002 and no signature of any persistent `flare' was observed.
The radio data clearly showed a tendency to go down from $1.0$Jy to $0.7$Jy reaching
at about $0.3$Jy on 28th/29th, while the X-ray data showed a
tendency to rise towards the end of the observation of the 27th.
The IR data in each band remained virtually constant. The H-band result was found to be higher
compared to the J and K$^\prime$ band results during 25th-29th September, 2002. A similar result
of turn around at about 4 micron was reported earlier by Fuchs (2003). This turn over could be
possibly due to free-free emission in optically thin limit.
Detailed discussion will be presented elsewhere (Chakrabarti et al. 2003).

In Fig. 2, we present the same light curves as in Fig. 1 but plotted around the `local' mean,
i.e., mean values taken in each `spell' of observation.
We note that there are significant variations in a matter of minutes in observations
at all the wavelengths. From eye-estimate, we see variability time-scale to be
$T_{var} \sim 2-8$ minutes. The error-bars include errors in individual measurements plus the
standard deviation of the flux variation in the light curve.
To impress that the variability is real, we show in Fig. 3 the
differential flux density variation of IR observations in the J and H
bands during 27 September 2002 using differential photometry. The error-bars are also 
shown. The 1 $\sigma$ error-bar (J=0.00035 Jy, H= 0.00085 Jy) of the differential flux
variation between SS433 and std1 for the whole light curve is a factor of
$3.5$ and $2.5$ in the J-band and H-band respectively in comparison to that 
between two standards (J=0.0001 Jy, H=0.00035 Jy). The 1 $\sigma$ for the
light curve is more than a factor of 5$\sigma$ of single point measurement
error. Thus, the variation in the IR light curves of SS433 is likely to be intrinsic
and the analysis shows above 2$\sigma$ level variability in both bands.

Could these variations be due to individual bullets? In order to be specific, we present in Fig. 4a,
one `micro-flare'-like event in radio from the data on 29th of Sept., 2002, when radio
intensity was further down $\sim 0.3$Jy so that the micro-flares could be prominently seen.
Here $0$s corresponds to $15$h$35$m UT. We observe brightening the source
from $0.35$Jy to $0.8$Jy in $\sim 75$s which faded away in another $\sim 75$s. That 
is, the intensity became more than doubled in $\sim 1$ minute! Similarly in Fig. 4b, 
where we presented a `micro-flare' from the 2nd (central) `spell' of X-ray data of 27th Sept. 2002 
(Fig. 1-2), we also observe significant brightening and fading in $\sim 100$s. 
Here $0$s corresponds to $16$h$5$m UT. The count rate went up more than $15\%$
or so in about a minute. The energy contained in the radio micro-flare, integrated 
over their lifetime is about  $I \nu \tau 4\pi D^2 10^{-23} = 1.1 \times 10^{33}$ergs
(Here, $I \sim 0.8$ is the intensity in Jansky, $\nu=1.28\times 10^9$Hz is the frequency 
of observation, $\tau\sim 100s$ is the rise-time of the bullet, $D=9\times 10^{21}cm$ 
is the distance of SS433). Similarly, the energy contained in the X-ray micro-flare is 
about $\frac{1}{2} \tau (N_{\gamma, max}-{\bar N_\gamma}) E_\gamma 4\pi D^2/A_{PCA}= 
2.7 \times 10^{35}$ergs (Here, $\tau \sim 100s$ is the rise-time of the flare,
$N_{\gamma, max}$ is the maximum photon count rate, ${\bar N}_\gamma$ is the 
average photon number, $E_\gamma$ is the average photon energy, $A_{PCA}$ is the
area of the PCA detectors.). The spectroscopic study
yields an average flux of $2.41 \times 10^{-10}$ ergs/cm$^2$/s. With an estimated 
duration of $100$s, and about $15\%$ energy going to the micro-flare (Fig. 4b), one obtains
the micro-flare energy to be $4.1 \times 10^{35}$ ergs in general agreement with 
the result obtained from photometric study. Since the radio luminosity is very small,
even when integrated over $0.1$ to $10$GHz radio band (with a spectral index of
$\sim -0.5$) (Vermeulen et al, 1993) we find that almost all the injected energy 
at X-ray band is lost on the way during its passage of $\sim 1-2$d. 

Though the variations we find are not periodic (strictly speaking they are
nor expected to be periodic, either), the power-density spectrum (PDS) does show
considerable power in frequencies $\sim 0.002-0.008$Hz.  
Deviation of the PDS from a power-law background 
$\propto \nu^{-\alpha}$ (e.g., Mineshige, Ouchi \& Nishimori, 1994) in all three 
bands gives an estimate of excess power at low-frequencies. 
We fit $\alpha=1.8$ for X-ray power, $1.9$ for IR power, and $1.6$
for radio power in PDS. X-ray power shows excess at $\sim 0.003$Hz 
($> 2.7 \sigma$), i.e., at  $T_{var, x} \sim 5.5$ min. and
at $0.0077$Hz ($ >1.4\sigma$), i.e., at $T_{var, x} \sim 2.1$ min. IR power shows excess 
at $0.0022$Hz ($>4\sigma$), i.e., $T_{var, ir} \sim 7.7$ min.
Radio PDS shows excess at $\sim 0.0023$Hz ($>3.2 \sigma$) 
i.e., at $T_{var, r} \sim 7.2$ min. and at $\sim 0.003$Hz ($>1.6\sigma$),
i.e. at $T_{var,r}\sim 5.5$ min. respectively. Here 1$\sigma$ error in residual power 
is the standard deviation  computed separately for each PDS after 
subtracting the power-law background $\nu^{-\alpha}$. Because the peaks in PDS are often marginally
significant we do not claim that we see quasi-periodic oscillations that are
observed in numerous black hole and neutron star candidates.

In order to establish that the features we observe are really due to `bullets' 
emitting at different wave bands, one should find correlations among them, or try to `follow' them
from one band to the other. Unfortunately, cross-correlation among our observations 
did not yield sharp peaks, partly because the observations were of short duration. 
Main problem is that the locations of the IR/radio emitting regions 
themselves are very uncertain. Also, the average duration of an `event' ($\sim$ minute) 
and average interval of the events ($2-8$ minutes) are very very short compared to 
the travel time of the bullets to IR ($\sim 10^{4}$s) or radio ($\sim 10^5$s) 
regions. However, we can exclude that the variabilities to be due to `fluctuations' at 
the inner regions of the accretion disk -- the typical time-scale of such variabilities 
(say, at $r\sim 3r_g$, where, $r_g=2GM/c^2$, $M$, $G$ and $c$ being the mass of the 
black hole, $G$ being the gravitational constant and $c$ being the velocity of light) 
of an $M=10M_\sun$ object would be $\sim 2\pi r/c \sim 20GM/c^3 \sim 10^{-3}$s, i.e., of 
much shorter duration than what we see.  Similarly, has it been due to 
random or flicker noise, we should have seen $1/f^{\alpha}$ ($\alpha \sim 1$)
dependence of the PDS. However, the best fit of PDS has $\alpha \sim 1.6-1.8$ instead. Thus,
the origin of these features must be different and could be due to bullet-like ejections from the disk.

In the spectrum, we find two strong Fe line features in all the three spells of X-ray observation.
The best fit was found to be the thermal bremsstrahlung model with two Fe lines
($kT \sim 18$keV) having a reduced $\chi^2$ of around $1.2$ in each case.
We failed to fit with a model having a blackbody emission component. Thus, no evidence for 
a Keplerian disk was found. The average flux was found to be $2.3 \times 10^{-10}$ ergs/cm$^2$/s.
This corresponds to a luminosity of $2.5 \times 10^{36}$ ergs/s. Since a bullet
has about $10-15\%$ of the total count (Fig. 4b), each bullet will have an energy 
of around $2.5 \times 10^{35}$ ergs/s.  

\section{Concluding remarks}

In this {\it Letter}, we presented results of our multi-wavelength observations which 
were save at short time intervals. From the analysis of the observations 
of radio, IR and X-ray in the quiescence state we conclude that we may be observing 
ejection events of bullet-like features from the accretion disk in time scales of $\sim 2-8$ minutes.
Identification of small micro-flare events 
with those those of bullet ejection is derived from the time scales of 
variabilities, which are roughly the same in all these wavelengths. 
We find their presence in X-ray ($\lsim 10^{11-12}$cm), IR ($\lsim 10^{13-14}$cm) 
and radio ($\lsim 10^{15}$cm) emission regions. Vermeulen et al. (1993) found 
evidence for optical bullets with life-time of $1-2$d. This is perhaps due to 
the propagation of a burst of indistinguishable bullets and not due to 
a single one. We identified micro-flare like features 
in all these observations which may be signatures of the
bullets. Count rate of X-ray was seen to increase $15-20$\% in a matter of a minute.
One way to actually identify each bullet could have been to follow them from X-ray region
outwards. This will require very careful time delay measurements since the 
distances of emission regions are not very accurately known to follow a
feature of duration of a minute. We exclude the 
possibility that what we see were flicker noise since neither the duration 
nor the PDS properties match with those of flicker noise. One could have 
perhaps distinguished the energetic bullets by observing polarization 
properties of the radio-emissions during the short-lived flares, but 
unfortunately due to technical reason this observation could 
not be carried out. We plan to do such an observation in near future.

SKC thanks Dr. J. Swank of NASA/GSFC and Prof. A. P. Rao of GMRT/NCRA for giving
time in RXTE and GMRT for observations. He also thanks Mr. J. Kodilkar
of GMRT/NCRA for his assistance in analysing the radio data. This work is supported 
in part by CSIR fellowship (SP) and a DST project (SKC and AN). Authors thank the
unknown referee for suggestions leading to considerable improvement of the observational aspects.

{}

\vfil\eject
\begin{figure}
\plotone{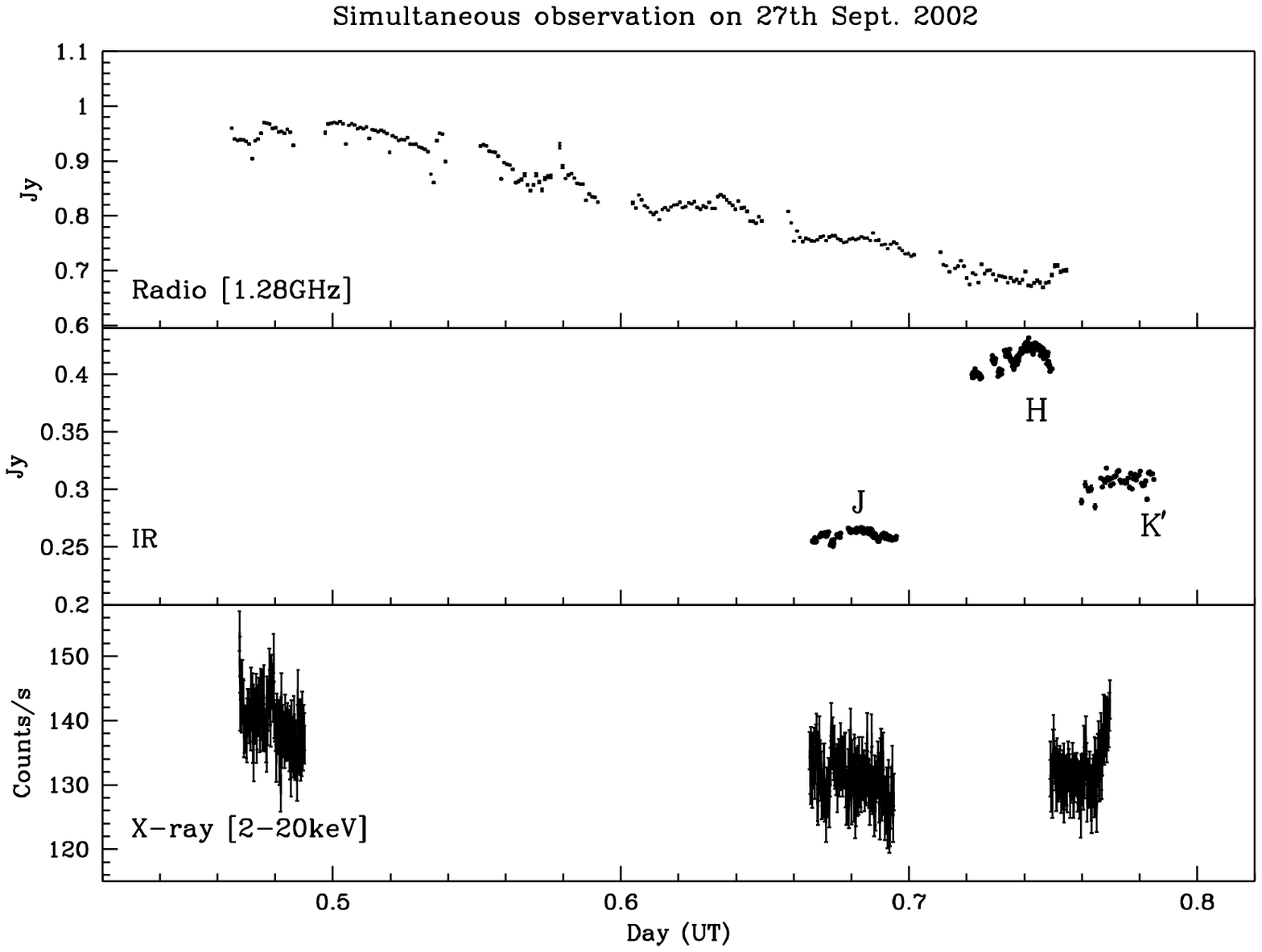}
\end{figure}

\vfil\eject
\begin{figure}
\plotone{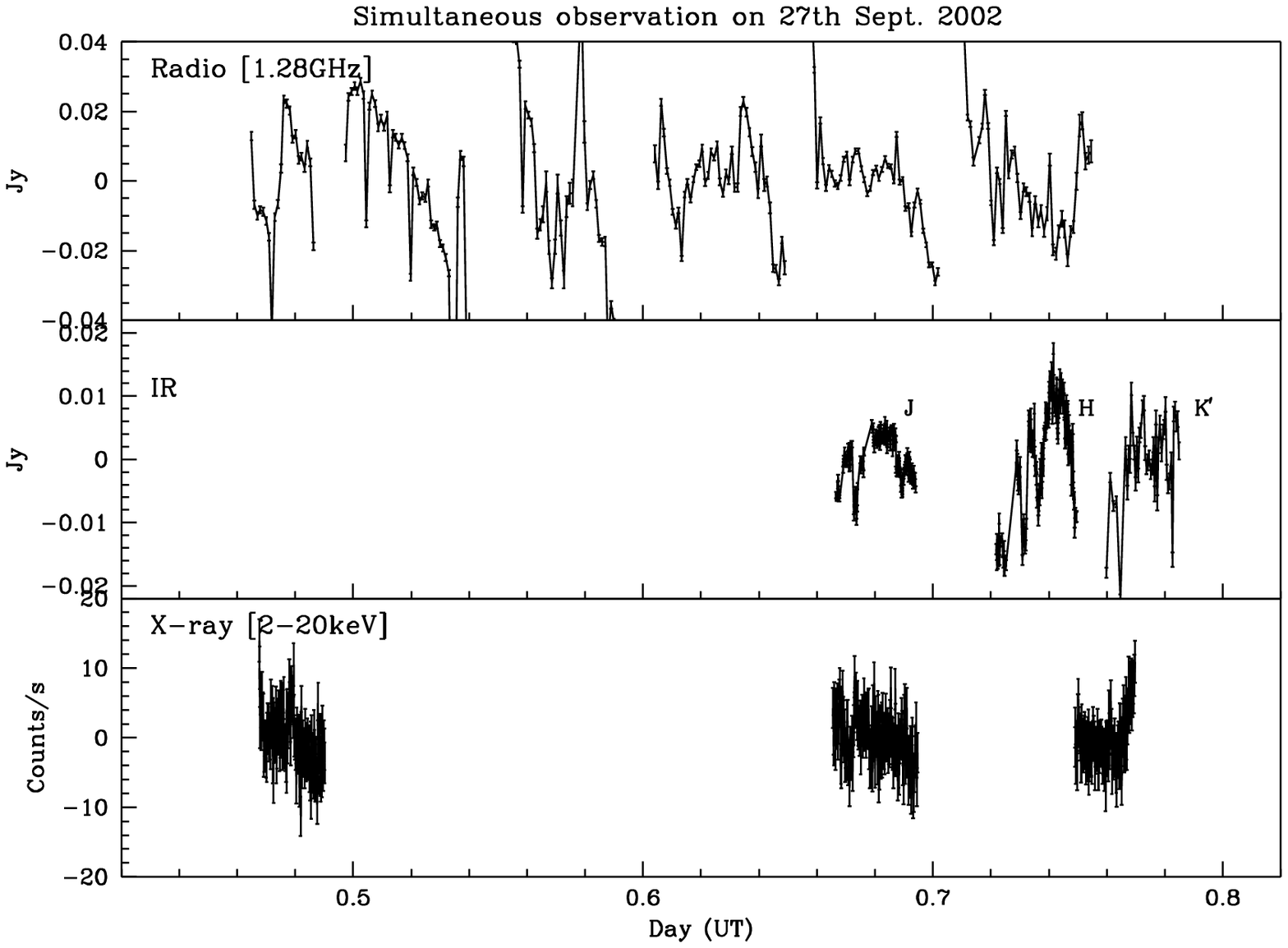}
\end{figure}

\vfil\eject
\begin{figure}
\plotone{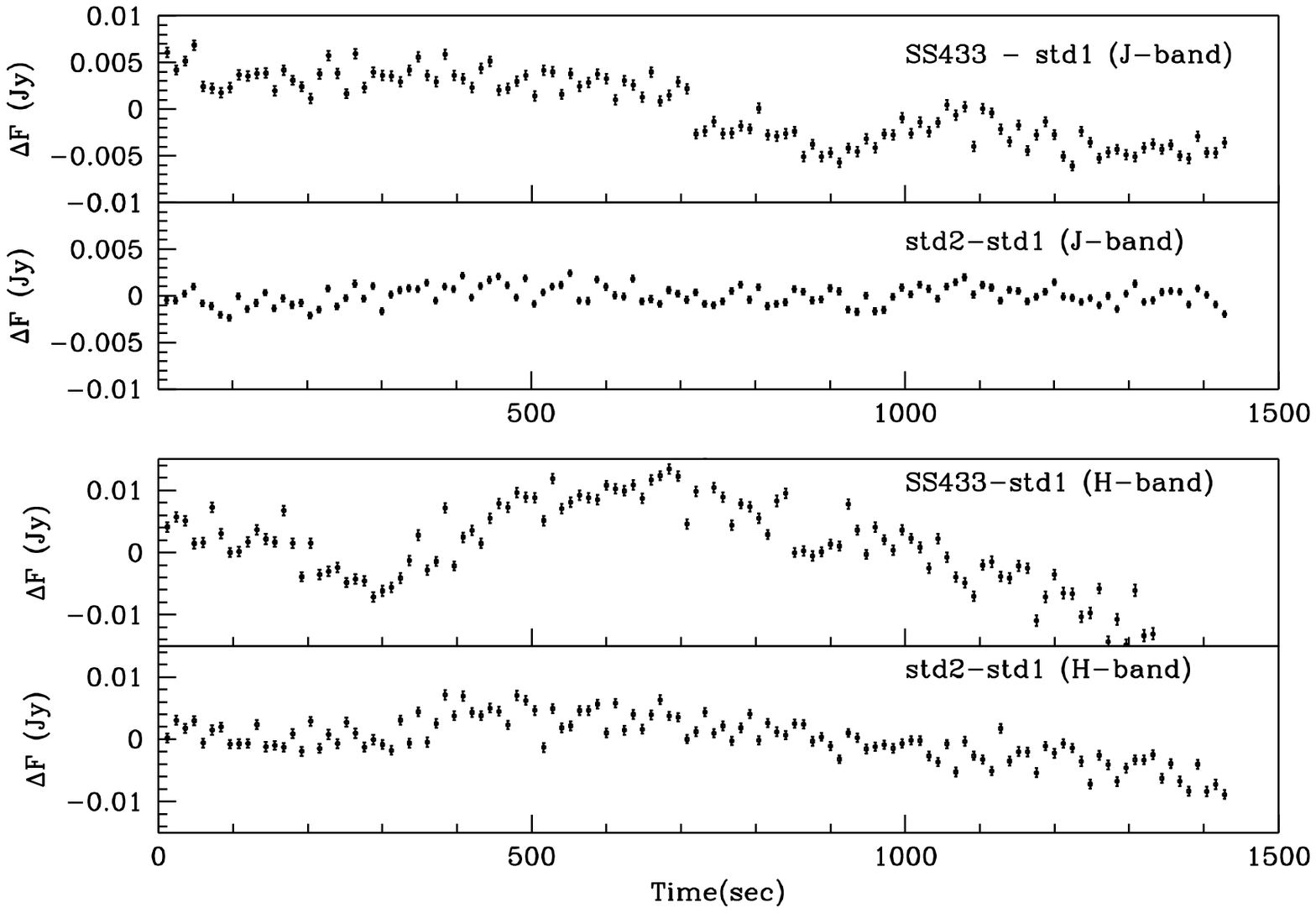}
\end{figure}

\vfil\eject
\begin{figure}
\plotone{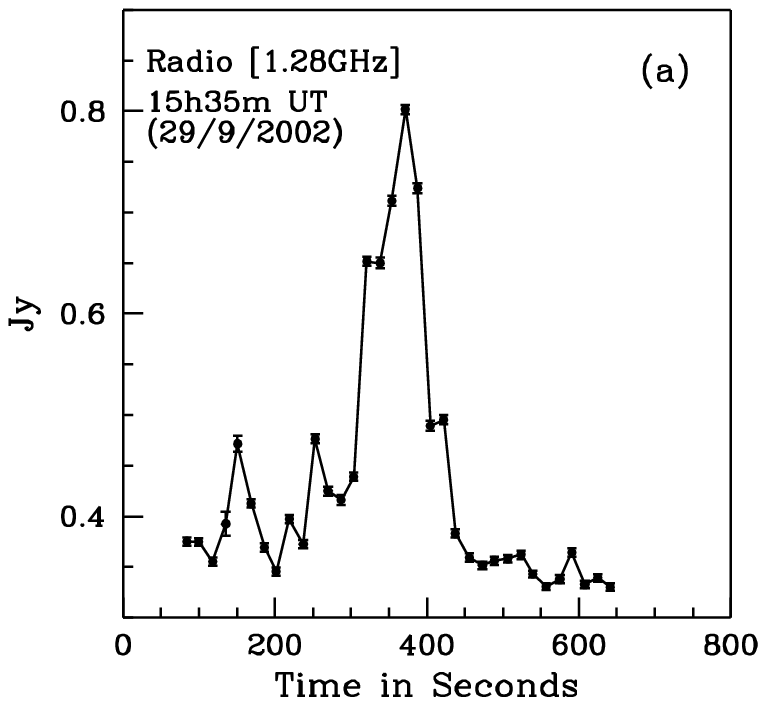}
\end{figure}

\vfil\eject
\begin{figure}
\plotone{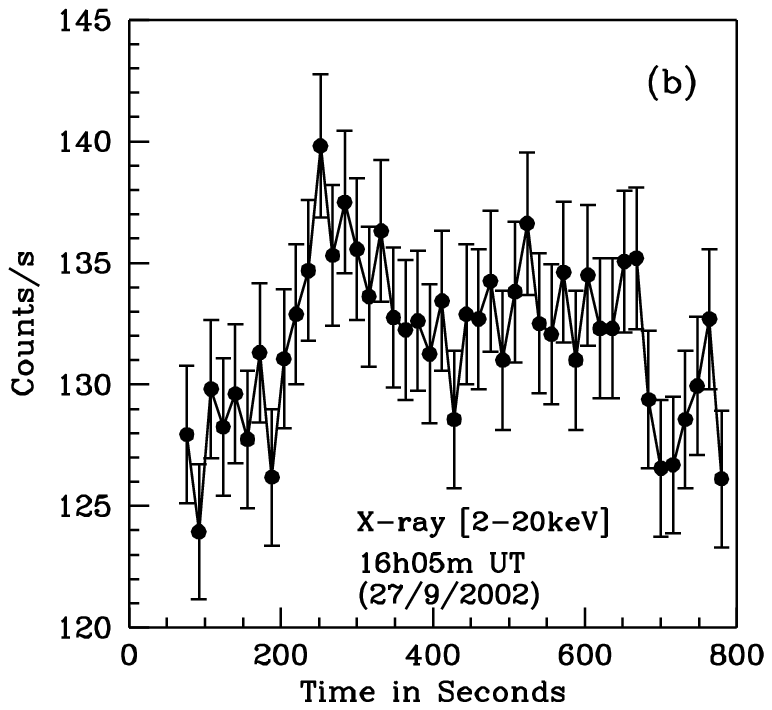}
\end{figure}

\newpage 

\noindent Fig. 1: Multi wavelength observation of short-time variability in SS433 by Radio (upper panel),
Infra-red (middle panel) and X-ray (lower panel) on 27th of September, 2002. The observations
were made at Giant Meter Radio Telescope, Pune at 1.28GHz (Radio), 1.2m Mt. Abu Infra-red
Telescope at J, H and K$^\prime$ bands and RXTE satellite (2-20keV) respectively.

\noindent Fig. 2: Observations at in Fig. 1 are plotted around mean taken in each spell
of observation. Considerable variations at time scale of a few minutes are observed.

\noindent Fig. 3: Differential photometry of SS433 with respect to two brightest standard stars
(std1 and std2) in  the same frame of the object are plotted. Different curves
are marked on upper-right corner in each panel. X-axis of the graph is the
relative time of measurements in seconds. The error bar for each individual
differential measurements are also shown. Differential flux variation of SS433
is above 2$\sigma$  level in comparison that of standards.

\noindent Fig. 4: Individual flares in very short timescales are caught.
(a) A radio flare lasting $2.5$ minutes (observed on 29th Sept. 2002)
and (b) an X-ray flare (observed on 27th of Sept. 2002) lasting for about $3.5$ minutes. Each bin-size is $16$s.

\end{document}